# Nano-thermodynamics of chemically induced graphene-diamond transformation

Sergey V. Erohin[1,2,3], Qiyuan Ruan[3], Pavel B. Sorokin[1,2], Boris I. Yakobson*[3]

[1] National University of Science and Technology MISiS, Moscow, 119049, Russia
[2] Technological Institute for Superhard and Novel Carbon Materials, Troitsk, Moscow 108840, Russia
[3] Department of Mechanical Engineering and Materials Science, and Department of Chemistry, Rice University, Houston, Texas 77005, USA

*Email: biy@rice.edu

Nearly two-dimensional diamond, or diamane, is coveted as ultrathin $sp^3$-carbon film with unique mechanics and electro-optics. The very thinness ($\sim h$) makes it possible for the surface chemistry, e.g. adsorbed atoms, to shift the bulk phase thermodynamics in favor of diamond, from multilayer graphene. Thermodynamic theory coupled with atomistic first principles computations predicts not only the reduction of required pressure ($p/p_\infty > 1 - h_0/h$), but also the nucleation barriers, definitive for the kinetic feasibility of diamane formation. Moreover, the optimal adsorbent chair-pattern on a bilayer graphene results in a cubic diamond lattice, while for thicker precursors the adsorbent boat-structure tends to produce hexagonal diamond (lonsdaleite), if graphene was in AA` stacking to start with. As adsorbents, H and F are conducive to diamond formation, while Cl appears sterically hindered.

Diamond is a material in no need of lengthy introductions, being highly coveted as precious gem, and for technologically important hard coatings, as a wide band gap semiconductor, for field-electron and also single-photon emission (SPE) from its certain color centers, N-V in particular. For all the above, with the exception of perhaps jewelry, the film-form is sufficient and may even be particularly attractive, which motivated a sustained effort in its fabrication for decades. Largely successful chemical vapor deposition[1–3] yields films with thickness from 20 nm to microns, but highly polycrystalline (5-10 nm grains), replete with grain boundaries and other structural defects. Growing a high quality diamond monocrystal-film remains an insurmountable challenge, possibly in the realm of sophisticated and costly high-pressure graphite-diamond transformation, also of rather limited area. To shift the thermodynamic balance from the $sp^2$-graphite to $sp^3$-diamond, one major driving factor has been the pressure of extreme values around 5-10 GPa. Moreover, there is also large kinetic barrier for diamond-graphite changes, in either direction. Even downhill spontaneous transformation from diamond to thermodynamically favored graphite ($\delta g \sim 20$ meV/atom lower) is inhibited by large barrier and would take geological times, good cause for a saying "diamonds are forever".[4] Nevertheless, at the nanoscale such diamond graphitization[5,6] does occur through the outer atomic layers. This process can be suppressed by saturation of dangling $sp^3$ surface bonds with adatoms or covalent functional groups, e.g. by hydrogenation, which "seals" the carbon in its energetically upper state, diamond. One could speculate that, conversely to graphitization, functionalizing the graphite surface can transform it to $sp^3$ state of diamond, to some depth; this however is hindered by the energy taxing $sp^3$-$sp^2$ interface, created underneath. Moreover, although the 2D-surface can affect many properties of 3D bulk material, obviously the surface state (reconstruction or chemical passivation) cannot change the thermodynamics of phase preference across entire volume. High pressure, assisted by temperature, remains prerequisite for getting 3D-diamonds from graphite.

With the advent of two-dimensional (2D) materials and particularly graphene (Gr), including its bilayer (BLG[7]) and few-layer (FLG[8]) varieties, this paradigm may change. In contrast to 3D bulk, if the sample is of very small, nanometer scale thickness, then its surface chemistry can switch the lattice organization (phase state) throughout. To appreciate the ease of such "phase conversion" by chemistry one recalls best studied monolayer graphene hydrogenated on both sides into CH composition. It was theoretically proposed[9] and christened "graphane" in its detailed study.[10] Basic notable features distinguishing graphane are the $sp^3$-hybridizaton of all C-atoms (instead of $sp^2$ in graphene) and its wide band gap (5.4 eV in the graphane chair conformation,[11] instead of zero in semimetal graphene), which justify considering it as ultimate, thinnest diamond slab[12] (especially since the bulk diamond surface is also typically H-passivated). The contrast in electronics of graphene and graphane invites possibilities of direct chemical patterning of functional circuitry[13,14]. The choice of active atoms is not limited to H, but can also be fluorine (interesting due to high chemical activity in its attachment to the graphene[13,15]), or chlorine.[16]

The kinetics of such transformation was first analyzed in the context of hydrogen storage[17] and spillover[18] media, showing the nucleation barriers strongly depend on the gas conditions, and vanish in hydrogen plasma (atomic H). In experiment, such conversion of graphene into CH (graphane) through hydrogenation,[19] into $CO_x$ (graphene oxide, GO) by oxidation,[20] or into CF (fluorographene) by fluorination,[21–23] all has been reported, usually as reversible transformations.

Single layer CH, in spite of its main attributes similar to diamond, is merely too thin to afford other important diamond properties like mechanical hardness and important ability to host key lattice defects. It was noted however that the affinity of graphene to H-plasma is sufficient to cause the fusion, induced by hydrogenation, of two graphene layers as well—into so called diamane,[24] that is bilayer graphane (over a feasible barrier),[25] and possibly several layers of FLG into hydrogenated form with interesting electronic properties,[26,27,28] including hosting N-V defects.[29] The surface functionalization can compel the lattice change underneath, though only to a small depth. The generic bulk diamond-graphite phase diagram cannot describe the $sp^2$-graphene to $sp^3$-diamane transformation, since the latter requires to explicitly including graphene surface, whose relative contribution brings about the nanoscale thickness $h$ as new thermodynamic variable. From thermodynamics standpoint, broad range of possibilities, quantified and mapped versus number of layers $N$ ( ~ $h$/3.4 Å), temperature $T$ and pressure $p$ (assuming full, uniform functionalization of graphene surface), suggested a possible path towards ultrathin diamond film,[30] through the hydrogenation or fluorination of the FLG. This path, however has not been explored even in theory, in spite of motivating progress in experimental evidence of feasibility of such $sp^2$-$sp^3$ transformation into extremely thin, nano-films of diamane.[31–37]

While theory suggests that surface-induced transformation can thermodynamically be favored up to 10-30 layers (3-10 nm), experimental evidence is scarce and limited to bilayers (or 3-4 when assisted by metal substrate[33]). There are apparently more severe kinetic limitations, not revealed by the thermodynamic phase diagram.[30] Here we focus on the mechanism of transformation, that is the "reaction path" from FLG to diamane, the sequence of specific intermediate configurations and their energies—essentially, the process of nucleation of the $sp^3$- phase from the initial $sp^2$-layered precursor, from (multilayer) graphene to diamane (thinnest diamond). This way we determine the intermediate of the highest energy that is the nucleation barrier, paramount for the transformation within realistic time. Bond by bond tracking, allowed in atomistic simulation, tells what final structure of the diamane is and how does it depend on the external conditions and the stacking of initial graphene layers. *Ab initio* atomistic analysis reveals the specific difference between the response of BLG or FLG atomic structure to either hydrogenation, fluorination or chlorination. While the focus is mainly on chemistry driven transformation, we also evaluate how a modest external pressure $p$ enables diamond formation from the films thicker than BLG.

**Results and Discussion**

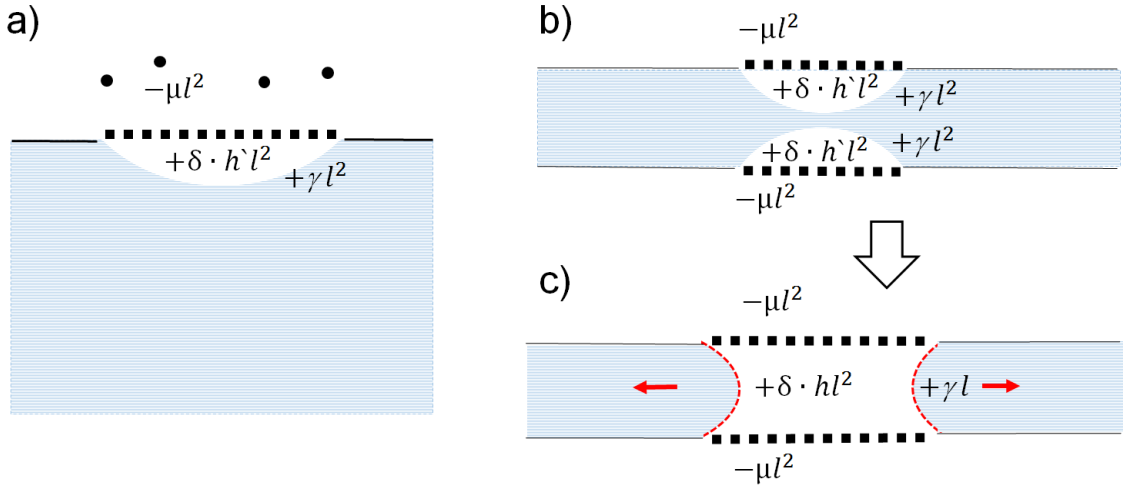

**Fig. 1.** Sketch of chemically induced phase transformation. Light blue region is the multilayer $sp^2$-graphene; thick-dashed segments mark the areas where active species (H, F, etc.) are chemisorbed to the newly-converted $sp^3$-diamane (white region). Red dashed lines are the $sp^3$-$sp^2$ interfaces, and the red arrows show the propagation of diamane phase.

It is helpful first to describe the process phenomenologically, in macroscopic, non-atomistic terms. By analytically approximating the free energy change $\Delta G(l)$ with the nucleus size $l$, in the course of diamane nucleation, one obtains an estimate of *thermodynamically* permitted thickness $h$, and even of the nucleation barrier $\Delta G^*$, defining if transformation is also *kinetically* feasible. A few energy contributions control the phase change, from the $sp^2$-graphite (or FLG) to diamond-like phase of $sp^3$-atoms. An increase ($\delta$, per unit) of internal energy is proportional to the volume of the phase-nucleus (Fig. 1a), characterized by its area $l^2$ times its depth $h$`: $\sim\delta \cdot h$` $\cdot l^2$. A volume reduction by $v$ gives a downhill term $-pv \cdot h$` $\cdot l^2$, if pressure $p$ is applied. Then the surface change cost is uphill in absence of passivation, but in presence of active adsorbents can be favorably negative, $\sigma$. It is a blend of gas atoms (H, F, Cl etc.) bonding to FLG surface, surface-bonds strain release, and possibly the reduction in gas-phase energy according to the chemical potential $\mu$ of its active reagent; all this contributes in proportion to the area $\sim \sigma \cdot l^2$ (we omit the numerical factors while concentrating on the dependencies and physical parameters). Another uphill term is due to the energy-costly interface between the $sp^3$-nucleus and surrounding $sp^2$-graphene multilayer, together with the misfit strain in the lattice around nucleus, and is proportional to nucleus interface area ~ $\gamma \cdot l^2$ (Fig. 1b), or ~ $\gamma \cdot h \cdot l$ – after the opposite nuclei already fuse in one (Fig. 1c).

Putting this together, in case of a bulk or a thick film (Fig. 1a), we arrive at $\Delta G(l) = (\sigma + \gamma + \delta \cdot h` – pv \cdot h`) \, l^2$. One sees that even favorably negative surface contribution $\sigma$ cannot overcome the uphill penalty terms, all of the same order $\sim l^2$; therefore, transition to diamond cannot occur in bulk, unless at very high pressure, $p > \delta/v$. In contrast, at small FLG thickness $h$, the opposite surfaces can reach across (Figs. 1b-c), the $\sim \gamma \cdot l^2$ terms "recombine" and vanish (a through-the-thickness $sp^3$-nucleus forms), and we arrive at different $\Delta G(l) = (\sigma + \delta \cdot h – pv \cdot h) \, l^2 + \gamma h \cdot l$. Now, the phase change is thermodynamically feasible if the first leading term is negative, at sufficient pressure

$$p > \delta/v + (\sigma/v)\frac{1}{h}. \qquad (1)$$

Two distinctly different cases arise from the surface conditions, defining the sign of $\sigma$. For transformation in vacuum, even with some reconstruction, diamond surface energy is higher than that of graphene, $\sigma > 0$. Then the pressure to induce diamond formation is larger for thinner FLG and increases rapidly for small h: the surface change is energetically taxing while its relative role is greater for fewer layers, so the greater $p$ is required.

More interesting is that with active sorbent $\sigma < 0$ and the needed pressure is lower for thinner films, becoming even negative, that is unnecessary ($p = 0$), for $h < h_0 \equiv |\sigma|/\delta$, estimated roughly as 10 nm or $N \sim 30$ layers of FLG, in ballpark agreement with *ab initio* thermodynamics assessment.[30] A dimensionless form is convenient, $p/p_\infty > 1 – h_0/h$, where $p_\infty$ is a pressure flipping the balance to diamond for infinite phase; for thin film it is easier. This is exactly the situation we aim to explore here, the chemically induced phase transformation. For a sample this thin, the penalty terms do not prevent the diamond phase spreading further (Fig. 1c), yet there is a nucleation barrier, the max of $\Delta G(l)$, now easily estimated as

$$\Delta G^* \approx (\gamma h)^2/(pv \cdot h + |\sigma| – \delta \cdot h). \qquad (2)$$

While not precise, this correctly captures the parametric dependencies: at $p = 0$, if thickness is near $h \sim \sigma/\delta$, the barrier rises to infinity, signifying thermodynamically prohibitive conditions. Moreover, for kinetic feasibility, this barrier must not exceed a few $k_bT$, which adds a criteria for thickness, $h < (k_bT \cdot \sigma)^{1/2}/\gamma$, estimating to a nanometer range—more stringent than sheer thermodynamics criteria above.

For analysis more informative quantitatively and even qualitatively (emergent lattice structure, its relation to the graphene stacking and surface hydrogenation patterns) we need next to consider explicit atomistic models, in form of a series of intermediate structures and their energies, computed from the first principles. Moreover, the continuum analytic relationships can then be fitted to these *ab initio* energies, so to determine the macroscopic parameters in the expressions and then use for a broader span of length scales, unaffordable for direct density functional theory (DFT) approach.

To this end, all atomistic structures under consideration were calculated by the DFT, within the Perdew−Burke−Ernzerhof generalized gradient approximation (PBE-GGA) [38] implemented in Vienna *ab initio* simulation package (VASP) [39–41]. The projector augmented wave (PAW) method [42,43] and the plane-wave cutoff energy of 400 eV were used. Grimme DFT-*D*3 corrections [44,45] with values of the dispersion coefficients [46] were applied to include van der Waals interactions of the layers. A vacuum size of 10 Å was set along *c* axis, perpendicular to the graphene planes, to simulate 2D structures within the periodic cell, containing the FLG-film as non-periodic cluster of 20–30 Å diameter, with H-passivated edges. Structure relaxation continued until the forces on atoms became less than 0.01 eV/Å. We mimic external pressure by a constant force along *c*-axis acting on each exterior atom in a region, through a code implementation in VASP. The corresponding Lagrange multiplier term is added to the total energy of the structure when performing minimization. Only atoms of outer layers (including cluster of adsorbed H) were placed in the region with force, and to specify pressure 10 GPa the force was set to ~0.15 eV/Å for each exterior atom.

Atomistic picture reveals immediately that the chemically induced phase transition occurs in specifically different ways for the single-, bi- and multilayer graphene cases. It can be seen directly by considering the behavior of chemisorbed H atoms on graphene films with various numbers of layers ($N$) in Fig. 2. Among the several possible ways to quantify the energy evolution with the degree of hydrogenation, we compute the average binding energy, $\varepsilon_b$, of the $n$ chemisorbed H-atoms as $\varepsilon_b = (E_{gr} + n\varepsilon_H - E_{nH@gr})/n$;[17] here, $E_{gr}$, $\varepsilon_H$ and $E_{nH@gr}$ are the energies of pristine graphene, single hydrogen atom, and graphene with $n$ attached H-atoms, respectively. The value of $\varepsilon_b$ tells if the gross reaction is favorable ($\varepsilon_b > 0$) or not ($\varepsilon_b < 0$). In the bilayer graphene case, the second ($n = 2$) atom H is attached in a counter, across the bilayer, energy favorable position, with the corresponding carbon atoms bonded, as H-C-C-H. This single bond is however only metastable, the energy is lower without it (transition from the lowest data-point in Fig. 2 vertically up, at $n = 2$). The actual layer bonding is favored at $n \geq 4$, as monotonous increase in average $\varepsilon_b$ begins.

Monolayer graphene hydrogenation displays a character of nucleation: although single H binding to graphene is weak, it strengthens dramatically as a compact graphane nucleus begins to form (Fig. 2, empty squares, topmost curve).[17, 25] The energy values, computed for gradual hydrogenation (increasing $n$) by adding H-atoms to both top and bottom surfaces of mono- or bilayer, display no nucleation barrier (assuming H source is in free atom form), distinguishing these cases from the thicker films.

Nevertheless, even in the two-layer case the initial adsorption of H atoms does not lead to connection of layers and formation of "diamond" nucleus. The binding energy of the first two atoms adsorbed on the BLG with C-C bond formed between the layers is smaller than C-H bonding of isolated layers by 0.25 eV. However, adsorption of two next H atoms already begins stably-bounded, energetically preferred diamond nano-nucleus. The favorable atomic structure of forming film is cubic diamond, whereas other possible diamond structures are higher in energy and less stable. However, such almost immediate bonding of the layers during hydrogenation also distinguishes BLG as a particular case, different from thicker, at $N > 2$.

Indeed, for the film of more than two layers the opposite surfaces do not "communicate" easily, so the hydrogenation starts from each one side without immediate bonding to underneath layers whose stable π-system does not engage unless by a sufficiently chemically-active $sp^3$ region of the preceding neighbor-layer. Herein, energy favorable conformation of semi-hydrogenated single layer (which imitates the hydrogenated multilayer graphene without bounded layers) is not chair- but boat-type [47] (called rectangular graphone). An expansion of $sp^3$-hybridized, in boat conformation, region leads to the increase of C-H binding strength, whereas the chair conformer becomes metastable and less favorable with every new adatom (red lowest curve in Fig. 2).

Eventually, these boat conformers of semi-hydrogenated graphene, at the opposite faces of the FLG, may connect through the rest of graphene layers, forming a new type of diamond film, hexagonal diamond (lonsdaleite). As we have shown previously, [48] such film can be formed only if the initial FLG has less-preferred AA' stacking and, therefore is not possible in the cases of AB (Bernal) or ABC ones. (One cannot exclude that at elevated temperatures, in the course of global annealing, either the H-atoms adsorption pattern can adjust from boat to chair, to enable the cubic diamond, or even the chemistry can drive the shear of weak-bonded layers into required stacking, to form lonsdaleite). These mostly-structural conclusions are not specific for hydrogenation process but extend to other possible active species of atoms, e.g. F and Cl.

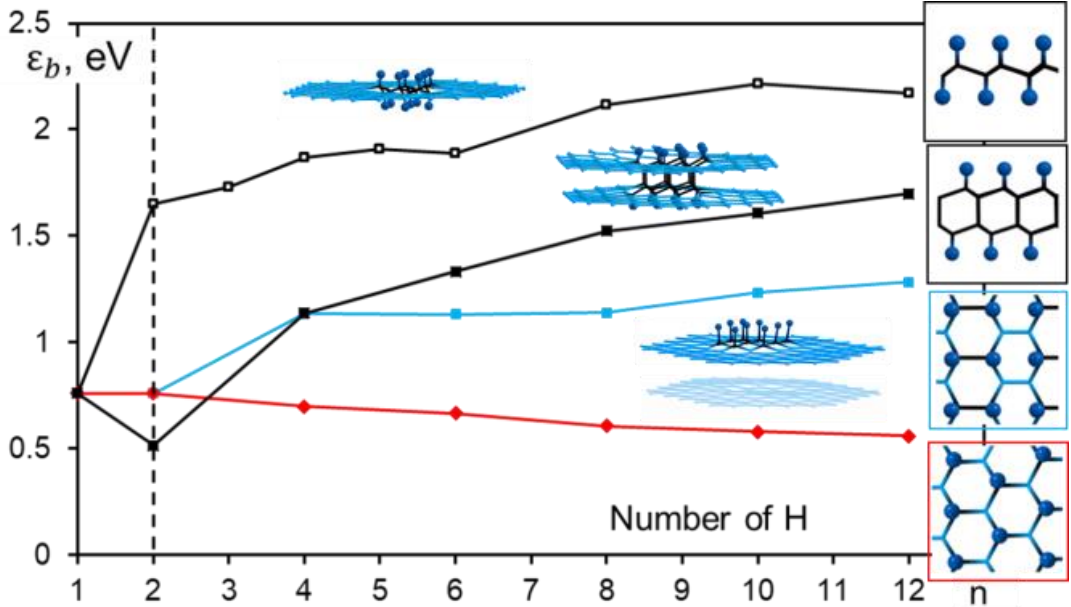

**Fig. 2**. Average binding energy, per H atom, in a two-side hydrogenated mono- (empty squares) and bi-layer (black solid squares) graphene, as well as graphene one-side hydrogenated, either in chair (red solid rhombs) or in boat (blue solid squares) conformations. (In the latter cases, the number $n$ of H atoms is doubled, for formal comparison with bilayer two-side hydrogenation.) The right-side insets show the atomic structures corresponding to each curve (blue or black bonds mark the $sp^2$ or $sp^3$ hybridized atoms).

The nucleation barrier of the diamond phase formation can be determined by tracking the Gibbs free energy as a function of nucleus size, $n$. For FLG with $n$ chemisorbed chemical species on both surfaces, the transformation energy $\Delta G$ is between the $sp^3$-diamond state of chemically bonded interior and the all $sp^2$-graphene layers,

$$\Delta G = \Delta E^{bulk}(N-2)n + \Delta E^{surf}n + \gamma + p\Delta V, \qquad (3)$$

it is explicitly decomposed into bulk, surface, interface and enthalpy term $pV$, if pressure is applied (details of the individual terms can be found in SI).

The first term, $\Delta E^{bulk}$, generally positive (~20-30 meV/atom), represents the difference in internal energy of carbon atom in diamond and graphite and, therefore, can be directly calculated from the energies of bulk lonsdaleite and graphite (we assume that amount of $sp^3$ atoms in each layer equals to $n$). This value is multiplied by internal layers' number, $N-2$.

The second is the negative surface-chemistry contribution, $\Delta E^{surf}$ being a difference in adatoms binding energy to surface layer bonded to the adjacent next-layers minus that for free unbonded layer. $\Delta E^{surf}$ term implicitly includes the strain in the surface layer due to its tendency to buckle at one-side functionalization [49] and is somewhat different for various adatoms. For H, the buckling is almost zero and, therefore, this value is very close to the (normalized per chemisorbed H) energy difference between hydrogenated bilayer diamond film and single flat graphene layer functionalized only from one side. For F, in contrast, one must add certain refinements, since a partial-local fluorination causes a high curvature of functionalized area. An external pressure $p$ can prevent such corrugation, keeping fluorinated layer flat. To account for this, the $\Delta E^{surf}(p)$ dependence on $p$ was approximated by a monotonous function satisfying both limits (SI): zero-pressure buckling and flattening under high pressure, similar to the mentioned hydrogenated film, as well as fitted to a few values $p = 3, 6, 10$ GPa.

The third term, $\gamma$, includes the energy of $sp^3$-nucleus interface with the multilayer $sp^2$-graphene it is imbedded in, and the strain of the latter caused by the denser diamond nucleus. The $sp^3$-$sp^2$ interface is essentially the same as for single graphene[17] and can thus be extracted from the data in Fig. 2 (top curve, empty squares). That plot follows the equation $\varepsilon_b \cdot n = \varepsilon_{\infty b} \cdot n - \gamma'\sqrt{n}$, consisting of the "bulk" contribution, $\varepsilon_{\infty b}$, proportional to a nucleus area (i.e. the number of H atoms, $n$), and the $sp^3$-$sp^2$ interface term proportional to number of interface atoms $\sqrt{n}$ with coefficient $\gamma' = 1.01$ eV. [17] For the present case of multilayer graphene it is multiplied by the number of internal layers, giving $\gamma'(N-2)$. Another part of the $\gamma$, the strain of FLG around the denser $sp^3$-nucleus, is due to the difference in the $c$ lattice constants of graphite and hexagonal diamond. Elastic energy caused by this mismatch one can estimate within continuum elasticity, so that interface contribution to the Eq. (3) amounts to

$$\gamma = \gamma'(N-2)\sqrt{n} + \frac{1}{2}C \iint h\varepsilon^2 dx dl, \qquad (4)$$

where $h = 3.4 \cdot N$ Å is a film thickness, $C$ is the elastic modulus of graphite in $c$ direction, $x$ is the distance from the edge of diamond nucleus, $l$ is a perimeter of strained area and $\varepsilon$ here is a strain. According to Saint-Venant principle, it must decay fast with distance, so that $\varepsilon \sim (\varepsilon_m - \varepsilon_p) \cdot \exp(-x/h)$, where $\varepsilon_m$ is the relative mismatch of hexagonal diamond and graphite lattices in $c$-direction, $\varepsilon_m = 0.37$ (cf. their densities, 3.5 and 2.2). To accommodate for external pressure $p$, the $\varepsilon_p$ is an inverse function of the dependence $p(\varepsilon_p) = 0.19\varepsilon_p + 4.99\varepsilon_p^2 + 4.36\varepsilon_p^3$, obtained from DFT calculations of graphite compression (SI). After integration the second term of Eq. (4) becomes $\gamma^G(p)N^2\sqrt{n} = 2.23(0.37 - \varepsilon_p)^2 N^2 \sqrt{n}$.

In addition to the last enthalpy term in (3) there is another pressure-induced contribution due to elastic compression energy, additive to the internal. Since the final, diamond state is nearly incompressible one can simply subtract the energy of the initial, graphene state, $\sim \frac{1}{2}p^2 V^G/C$, or more precisely $\sim \frac{1}{2}\varepsilon_p p V^G$, if the nonlinear function $\varepsilon_p$ is known; the volume $V^G$ is a nominal, free graphene volume under $sp^3$-hybridized surface area $A$: $V^G = hA = 8.3Nn$ Å$^3$ (8.3 Å$^3$ per C-atom in graphite).

Adding the above together, the expression for $\Delta G$ is:

$$\Delta G = \Delta E^{bulk}(N-2)n + \Delta E^{surf}n + \gamma'(N-2)\sqrt{n} + \gamma^G(\varepsilon_m - \varepsilon_p)^2 N^2 \sqrt{n} - \frac{1}{2}\varepsilon_p p V^G - (\varepsilon_m - \varepsilon_p)pV^G. \qquad (5)$$

Here, the surface-chemical term $\Delta E^{surf}$ (do not scale with thickness $N$) and pressure-driven terms ($\sim N$) are negative, representing the general trend of phase transition induced by *surface* chemisorption and external *pressure*. The uphill, positive terms all scale as $\sim N$ or even $\sim N^2$, and for large number of layers, $\Delta E^{bulk}$ term (phase-change energy, $\sim N$) and costly mismatch strain ($\sim\gamma^G N^2$) overcome the surface chemistry gains and thus prohibit the nucleation. For thicker FLG, both critical size of $sp^3$-cluster and nucleation barrier increase rapidly. All coefficients to Eq. (5) appear in the Table 1 below.

|   | $\Delta E^{bulk}$, eV | $\Delta E^{surf}$, eV | $\gamma'$, eV | $\gamma^G$, eV | $V^G$, Å$^3$ |
|---|---|---|---|---|---|
| H | 0.043 | -0.85 | 1.01 | 2.23 | 8.3 $Nn$ |
| F |  | $\Delta E^{surf}(p)$ |  |  |  |

**Table 1.** Numerical values of the parameters in the Eq. (5).

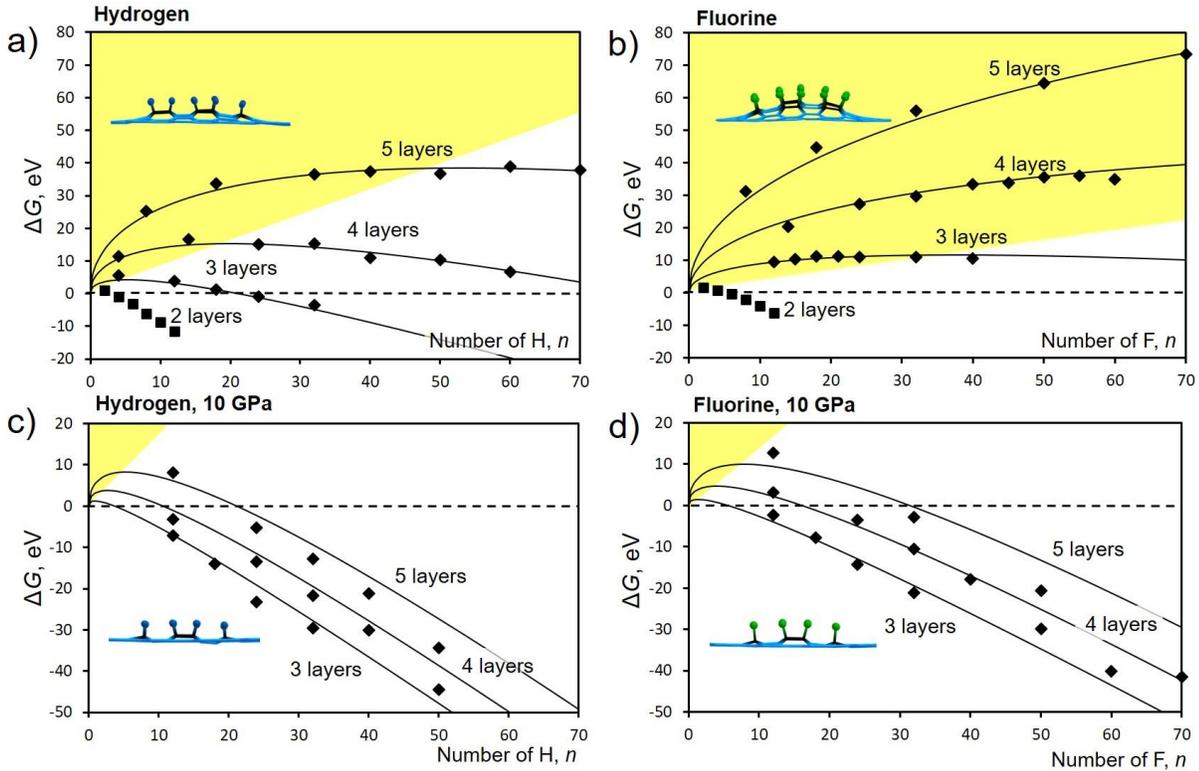

**Fig. 3.** $\Delta G$ plotted versus number of H (hydrogen, a,c) or F (fluorine, b, d) atoms chemisorbed at the surface, without external pressure $p = 0$ (a, b) and at $p = 10$ GPa (c, d). Points are from direct DFT calculations, while solid lines represent Eq. (5). The insets show the difference between buckling amplitude for hydrogenated and fluorinated monolayer for $p = 0$, and $p = 10$ GPa. Yellow shading marks the area of diamond nucleus instability and reversal, preventing diamane formation.

The Fig. 3 combines the energies $\Delta G$ from *ab initio* computed (relaxed) metastable configurations of the FLG with different number of chemisorbed H or F atoms (diamond symbols), in good agreement with the Eq. (5) curves. The negative values mean the nucleation can be overall favorable. For the large structures with $n > 70$ (of H or F) direct DFT is unaffordable. Nevertheless, calculated data clearly show the trends, and estimate the number of adatoms required to achieve a chemically induced phase transition. The extrema of the curves indicate the successful nucleation when further chemisorption of H/F proceeds as exothermic growth of diamane phase, and the extrema height is the nucleation barrier.

From Fig. 3 plots, one can see that the nucleation barriers for fluorination process are surprisingly higher than for the hydrogenation. Moreover, for the chlorine case simulations reveal that the ordered boat-conformation on the surface is entirely unstable. This is apparently due to the atomic radii of adatoms crowded in the compact cluster and causing more initial buckling of the surface layer (compare atomic structure of semi-hydrogenated and semi-fluorinated regions in the insets of Fig. 3a-b). The convex shape impedes the bonding of layers.

Nevertheless, even in the case of H, the barriers for chemically induced phase switch are high, and an external pressure can be an obvious remedy. We evaluate the effect of $p = 10$ GPa, readily achievable experimentally.[34] Compressing film by ~10% notably changes the elastic contribution (represented by $\sim pV^G$ terms) and decreases the nucleation barriers, Fig. 3c-d. The Eq. (5) and even its simple version (2) readily show how the nucleation barrier and critical nucleus size depend on the external pressure $p$ and on the film thickness $N$, plotted in Fig. 4. While at $p = 0$, the barrier for F-case is much higher, with increasing pressure it falls faster and reaches that for H. If $p$ suppresses the surface buckling, the behavior of fluorinated and hydrogenated films become similar. At $p = 10$ GPa both hydrogenated and fluorinated three-layer graphene films must overcome ~2 eV barrier, for the nucleus 4-6 adatoms, a very reasonable condition. Note that overall behavior of the $\Delta G^*(p)$ plots agree well with Eq. (2).

It may be instructive comparing predicted barriers with available for the bulk diamond phase formation. Simulations of homogeneous diamond nucleation in graphite at 20 GPa encounters 560-630 eV barrier,[50] hundreds times higher than what we predict for 3-4 layer diamane formation (independent of adatoms type). In practice, the insurmountable barriers for the bulk diamond nucleation decline by heterogeneous nucleation in presence of graphite structural defects, whereas we predict possibility in principle to produce diamane from graphene without defects, a great benefit for fabricating perfect diamond nano-films for future electronic devices.

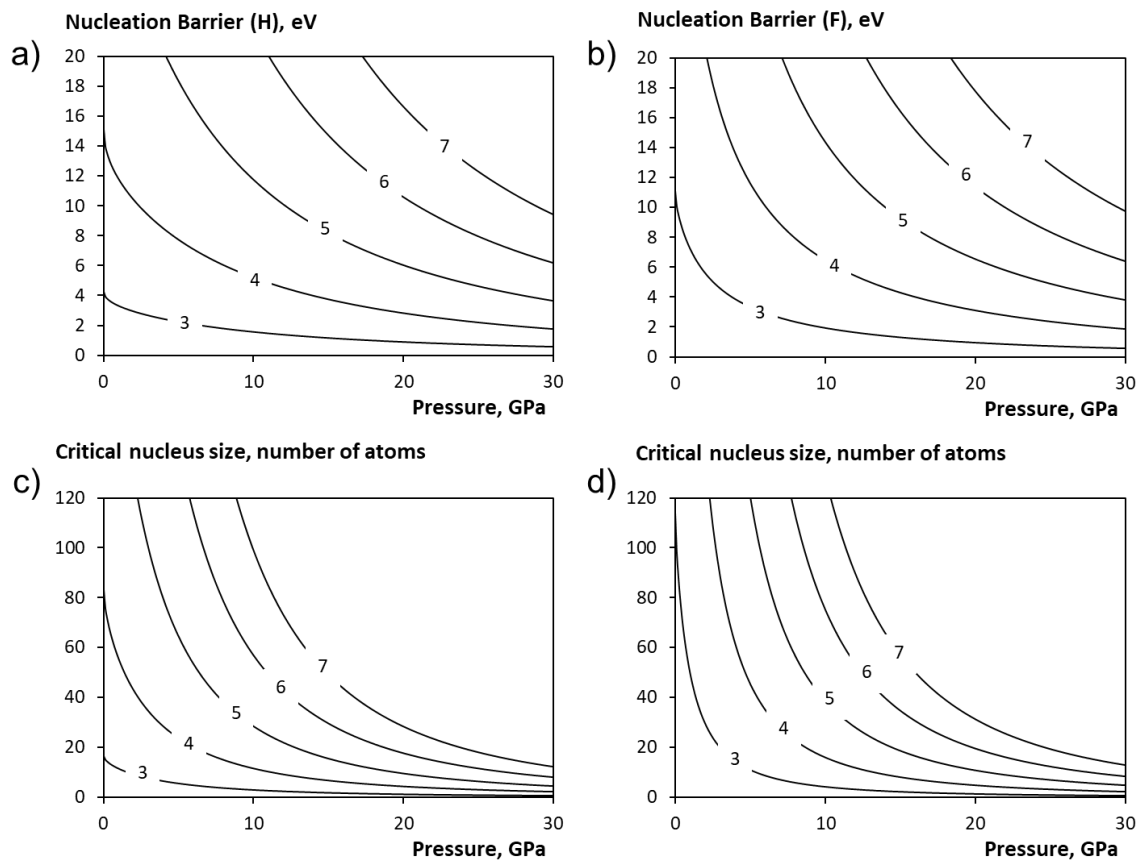

**Fig. 4.** The computed nucleation barrier (a, b) and nucleus size (c, d) versus pressure for H- (a, c) and F-atoms (b, d) chemisorption, for different number of layers (*N*, integers on the curves).

We do not explicitly consider the initial gas form. $H_2$ and $F_2$ dissociation is endothermic, by 2.3 eV [17] and 0.97 eV [51] per atom, respectively, suggesting that atomic H of F form (cold plasma) should be used, or perhaps $XeF_2$ (xenon difluoride) as a known powerful fluorinating agent, as recently shown to work well.[37] From chemisorbed stage forward, since for $N > 2$ the chemisorbed energy is lower for preferred boat-pattern (Fig. 2, inset), the chemically induced phase transition yields diamane films of hexagonal structure (notably, bulk single crystal lonsdaleite has not been achieved up to date). The observation that the thicker films are expected to form with lonsdaleite structure, while bilayer turns rather into cubic diamond, is intriguing enough to deserve further attention, especially for possibly discriminating in experiments. Altogether, Fig. 4 plots show rather high nucleation barrier, underlining the importance of its suppression, by combining pressure and chemistry, in order to form diamane of more than two layers. The precursor graphene stacking is another important parameter, often somewhat overlooked (without taking it into account, the diamondization may be expected to occur for only exterior layers.[52,53])

**Conclusions**

In summary, combining nanoscale thermodynamics theory with atomistic DFT simulations enables quantitative analysis of chemically-induced phase transformation, from multilayer graphene to diamane, a nanometer thin film of diamond. The main driving factor is the active species, exemplified here by H, F, or Cl atoms, chemisorbed to the exterior surfaces of multilayer graphene, and favoring transition into $sp^3$-state. This can of course be further assisted by external pressure, whose thermodynamic threshold appears significantly reduced for thin films, $p(h) = p_\infty \cdot (1 - h_0/h)$, relative to bulk phase thermodynamics value, $p_\infty$. More stringent, kinetic requirements are revealed from the diamond nucleation in multilayer film, establishing its barrier dependence on the film thickness $h$ (i.e. number of graphene layers $N$), on the active adsorbent-atoms (H, F, or Cl) and the assisting applied pressure, $p$. It is found that the diamond film formation in the case of bilayer and thicker can significantly differ: from bilayer graphene, diamane can form spontaneously, while the thicker ($N > 2$) films encounter a nucleation barrier, the more layers the higher, necessitating pressure assistance. Computations show however (Fig. 4) that high pressure is only needed to create a small nucleus, and therefore in practical realization can be applied only locally (1-2 nm) and for rather short time (perhaps, milliseconds), a needle-prod. After that, the diamond film can grow either fully chemically-induced or under much milder thermodynamic pressure (above). Further, morphologically, a bilayer is likely to concurrently chemisorb atoms in a chair-pattern and produce cubic diamond structure. In contrast, films thicker than two layers opt to the (lower energy) boat-type chemisorption and are consequently expected to form hexagonal diamane film, a lonsdaleite; notably this seems to require AA' packed graphene, not readily available but can also possibly reshuffle in the course of chemically-driven transformation (the latter speculation may need to be looked at separately). The choice of active atoms also appears important, as discussed for H, F, and Cl. Overall, one can conclude that a combination of surface chemistry with properly chosen active species X (gas, atoms, or possible direct metallic contacts) with moderate pressure regime look as promising avenue to fabricating nanoscale diamond films by conversion from few-layer graphene, symbolically X + C$sp^2$ → XC$sp^3$.

## ACKNOWLEDGMENTS

S.V.E. and P.B.S. were supported by Ministry of Education and Science of the Russian Federation in the framework of Increase Competitiveness Program of NUST MISiS (K2-2019-016), with computational resources of Materials Modeling and Development Laboratory at NUST MISiS and Joint Supercomputer Center of the Russian Academy of Sciences. P.B.S. acknowledges Grant of President of Russian Federation for the young DSc. (MD-1046.2019.2). Work at Rice (S.V.E., Q.R., B.I.Y.) was supported by the ARL-Rice Initiative and the Office of Naval Research, grant N00014-19-1-2191.